\documentclass[aps,prd,secnumarabic,amssymb,nobibnotes, amsmath,twocolumn,nofootinbib,superscriptaddress,floatfix,preprintnumbers]{revtex4-1}

\usepackage{graphicx,color}% Include figure files
\usepackage{dcolumn}% Align table columns on decimal point
\usepackage{bm}% bold math
\usepackage{lineno, soul, ulem}
\usepackage[unicode=true]{hyperref}% add hypertext capabilities
\usepackage{amsmath}
\usepackage{mathtools}
\usepackage{multirow}

\usepackage{aas_macros}
\usepackage{xcolor}
\newcommand{\spi}{SPI}
\def\d{{\rm d}}

\def \aap  {A\&A}

\def \apjs  {ApJS}
\def \apjl  {ApJL}

\def \jcap  {JCAP}
\def \prd {Phy. Rev. D}

\def \mnras {MNRAS}

\begin{document}

\begin{flushleft}
LAPTH-006/22
\end{flushleft}

\title{Strong constraints on primordial black hole dark matter from 16 years of INTEGRAL/SPI observations
}

\author{J.~Berteaud}
\email{berteaud@lapth.cnrs.fr}
\affiliation{LAPTh,  CNRS, USMB, F-74940 Annecy, France}
\author{F.~Calore}
\email{calore@lapth.cnrs.fr}
\affiliation{LAPTh,  CNRS, USMB,  F-74940 Annecy, France}
\author{J.~Iguaz}
\email{iguaz@lapth.cnrs.fr}
\affiliation{LAPTh, CNRS,  USMB, F-74940 Annecy, France}
\author{P.~D.~Serpico}
\email{serpico@lapth.cnrs.fr}
\affiliation{LAPTh, CNRS, USMB,  F-74940 Annecy, France}
\author{T.~Siegert}
\email{tho.siegert@gmail.com}
\affiliation{Institut f\"ur Theoretische Physik und Astrophysik, Universit\"at W\"urzburg, Campus Hubland Nord, Emil-Fischer-Str. 31, 97074 W\"urzburg, Germany}
\affiliation{Max Planck Institut f\"ur extraterrestrische Physik, Gie\ss enbachstr. 1, 85748, Garching bei M\"unchen, Germany}

\begin{abstract}
We present a new analysis of the diffuse soft 
$\gamma$-ray emission towards the inner Galaxy as measured 
by the SPectrometer aboard the INTEGRAL satellite (\spi) with
16 years of data taking.
The analysis implements a spatial template fit of \spi~data and an improved instrumental background model. 
We characterize the contribution of Primordial Black Holes (PBH)
as dark matter (DM) candidates evaporating into ${\cal O}$(1) MeV photons by including, for the first time, the spatial distribution of their signal into the fitting procedure.
No PBH signal is detected, and we set the strongest limit on PBH DM for masses up to $4 \times 10^{17}$ g, significantly closing in into the so-called asteroid mass range.
\end{abstract}

\maketitle

%%%%%%%%%%%%%%%%%%%%%%%%%%%%%%%%%%%%%%
\section{Introduction}
%%%%%%%%%%%%%%%%%%%%%%%%%%%%%%%%%%%%%%
A long-standing, paradigmatic non-particle candidate for the dark matter (DM) in the universe is constituted by Primordial Black Holes (PBHs) (see Ref.~\cite{Green:2020jor} for a recent review). Such objects could have arisen in the early universe from the gravitational collapse of overdensities made of {\it ordinary} radiation and/or matter, associated for instance to large density fluctuations at small scales set by non-minimal inflationary settings, or to phase transitions.  While most of the PBH DM parameter space is excluded or tightly constrained by a number of observations, a window of masses between about $10^{17}\, $g and $10^{23}\,$g, i.e.~the {\it  asteroid} mass range, is still potentially viable. The upper range of this window can be tested via high-cadence micro-lensing surveys, the lower part is accessible via high-energy astrophysical probes, sensitive to their Hawking evaporation spectrum, falling in the hard X-ray to the soft $\gamma$-ray band. Large fields-of-views and high statistics measurements in this observationally challenging window are required to tighten this probe. One instrument with appropriate characteristics is the coded-mask spectrometer telescope, SPI, aboard the INTEGRAL satellite \cite{Winkler2003_INTEGRAL,Vedrenne2003_SPI}. SPI surveys the $\gamma$-ray sky with a focus on the Galactic bulge and disk and, with a forward application of the imaging response,   {can} observe diffuse emission  such as expected from cosmic-ray interactions with gas and radiation fields or from the distribution of DM in the Milky Way.

We present here the first, dedicated, \spi~analysis of
PBH DM in the Milky Way: The spatial distribution of the PBH MeV signal is 
considered as an independent {\it template} in the fit of \spi~data, as
it is done for other known astrophysical components, such as the diffuse inverse Compton (IC) scattering along the Galactic plane, or the Positronium (Ps) emission with a strong $\gamma$-ray line at 511\,keV.
We go beyond previous approaches, in particular Ref.~\cite{Laha2020_PMBHDM},
where the limits on PBH DM are inferred from a measurement of the soft $\gamma$-ray diffuse emission derived with SPI data using a set of templates which do not account for a PBH component.
Consequently, the limits derived with such an approach may be biased, since the additional PBH contribution is neglected 
in the set of templates adopted. Note that the PBH evaporation emission has a specific {\it morphology}, following generic DM density profiles, e.g.~a Navarro-Frenk-White (NFW) profile~\cite{Navarro1997_NFW}, which must be taken into account to derive a fully self-consistent
limit on the PBH parameter space.
SPI analyses are challenging and require great carefulness: Besides diffuse astrophysical fluxes,  several hundreds of (variable) point source contribute to the total signal, especially at energies $\lesssim$ 300 keV.
This is particulary relevant for constraining higher PBH masses, since the PBH blackbody temperature decreases with mass as $T_{\rm PBH} \propto M_{\rm PBH}^{-1}$.

{This article is structured as follows: Sec.~\ref{sec:templatefit} describes our new \spi~template analysis. Sec.~\ref{sec:spfits} is a compact description of the spectral components entering the fit to the total spectrum.
Sec.~\ref{results} presents our results, while Sec.~\ref{discconcl} includes  a discussion and conclusions. Further technical information is provided in appendices: Appendix~\ref{sm:dataset} is  describes the dataset and its handling; Appendix~\ref{sm:sys_ic} is devoted to the template fitting procedure and systematic uncertainties, while appendix~\ref{sm:spectralfit} includes further details on the spectral fitting procedure and results, including astrophysical background components. }
%%%%%%%%%%%%%%%%%%%%%%%%%%%%%%%%%%%%%%
\section{New \spi~template analysis of diffuse $\gamma$ rays} 
\label{sec:templatefit}
%%%%%%%%%%%%%%%%%%%%%%%%%%%%%%%%%%%%%%
We analyzed 16 years of data taken by \spi~over the energy range
30\,keV--8\,MeV, extending to low energies the new measurement of the diffuse soft $\gamma$-ray emission with SPI between 0.5 and 8 MeV recently performed by some of the authors in Ref.~\cite{Siegert:2022jii}.
{The lower energy limit is such to avoid source confusion especially towards the Galactic Center, and to guarantee correct energy calibration and detector performance.}

Our analysis relies on an improved description of the instrumental 
background (BG) \cite{Diehl2018_BGRDB,Siegert2019_SPIBG}, and a systematic study of the dominating diffuse IC scattering emission which originates from cosmic-ray electrons in the GeV range produced by standard astrophysical sources in the Galactic disk.
Details on the dataset used can be found in Appendix~\ref{sm:dataset}.

Our \spi~data analysis relies on the comparison of models, i.e.~images, and a description of the instrumental BG, to data in a raw format, here the number of photons detected per unit observation per detector per energy bin.
The images (spatial templates or morphologies) are convolved through the \spi~coded-mask response, which depends on the source aspect angle and the photon energy.
The BG model contains all the knowledge about the instrument, the detectors, and their behaviour 
over the full \spi~data taking period.
The models can then be considered as time-series of expected detector patterns which are fitted to the measured time-series via a maximum likelihood approach, using the Poissonian likelihood. This is done energy bin by energy bin independently, to extract the spectrum of the template maps included and to obtain the total flux. So, unlike in the case of the {\it Fermi} Large Area Telescope (LAT), where the measured counts are projected onto a sky map and where the resulting image is interpreted in terms of different templates, \spi~can hardly provide morphology-independent spectra nor spectrum-independent morphologies with the current standard software, OSA/\textit{spimodfit}~\footnote{{User manual available at\\
http://isdc.unige.ch/integral/download/osa/doc/11.0/\\
spimodfit\_handbook.pdf}} \cite{Courvoisier03,Strong2005_gammaconti}. Instead, the spatial templates for one or more components are assumed and fitted to the raw data to obtain their flux contribution.
{Spectral fits on the components
that have been separated via angular templates
are performed in a second step, described in Sec.~\ref{sec:spfits}.}

% ROI
Considering the fully-coded SPI field of view of $16^{\circ} \times 16^\circ$
and the partially-coded (i.e.~without all detectors exposed to the source) field of view of $30^\circ \times 30^\circ$, we select a {\it region of interest} (ROI) in this analysis of  $|l| \leq 47.5^\circ$, $|b| \leq 47.5^\circ$. 
This is the range over which we calculate our spatial templates.

% Spi model
Besides the instrumental BG, \spi~data are fitted with a model for 
astrophysical sources of MeV photons. 
Below $\sim 300$\,keV, the total emission is dominated by point sources \cite{Bouchet2011_diffuseCR}. 
 {We assume that all source included in this study are constant in time. This means an average of the sources' fluxes is extracted. This can partially impact the diffuse emission components in particular below $\sim 50$\,keV.}
We adopt an iterative approach (similar to \cite{Strong2005_gammaconti}) to determine the contribution of point sources at each energy. This means we use the compiled \spi~source catalog from Ref.~\cite{Bouchet2011_diffuseCR}, including 256 known sources for the full sky, and fit the total model, i.e.~all diffuse components (see next paragraph) plus point sources plus instrumental BG, per energy bin. Unless a source is detected with more than $2\sigma$ in two subsequent energy bins, we drop the source to reduce the number of fitted parameters for the next-higher energy bin. 
 {The point sources positions are taken 
from~\cite{Bouchet2011_diffuseCR}, so that the resulting diffuse spectrum is indeed similar to the one in~\cite{Bouchet2011_diffuseCR}.
Besides, because of the source variability,  our method includes a certain higher degree of systematics compared to \cite{2013A&A...555A..52B}, but because
we know where the sources are statistical uncertainties are reduced.}
We find sources up to $\sim 1$\,MeV in our ROI.  {However, because of source confusion and the limits of our method to handle more than 10000 parameters, we get a reliable estimate of point source fluxes only above 50 keV.}

% total spectrum
To extract the total diffuse spectrum we consider the following {spatial} components: 1) A population of unresolved point sources up to $\sim 100$\,keV \cite{Krivonos2007_GRXE}; 2) the Ps emission including continuum up to 511\,keV and the 511\,keV line \cite{Siegert2016_511};  3) the $\mathrm{^{26}Al}$-line at 1809\,keV from massive stars \cite{Diehl2006_26Al}; 4) the diffuse IC scattering continuum in the whole spectral range \cite{Siegert:2022jii}; 5) the expected morphology from evaporating PBHs, tracking the conventional NFW halo for DM distribution, detailed in the following;
{see also Fig.~\ref{fig:diffuse_models} in Appendix~\ref{sm:sys_ic}.
}
\begin{figure}[t]
\includegraphics[width=0.49\textwidth,trim=0.0in 0.1in 0.0in 0.1in,clip=True]{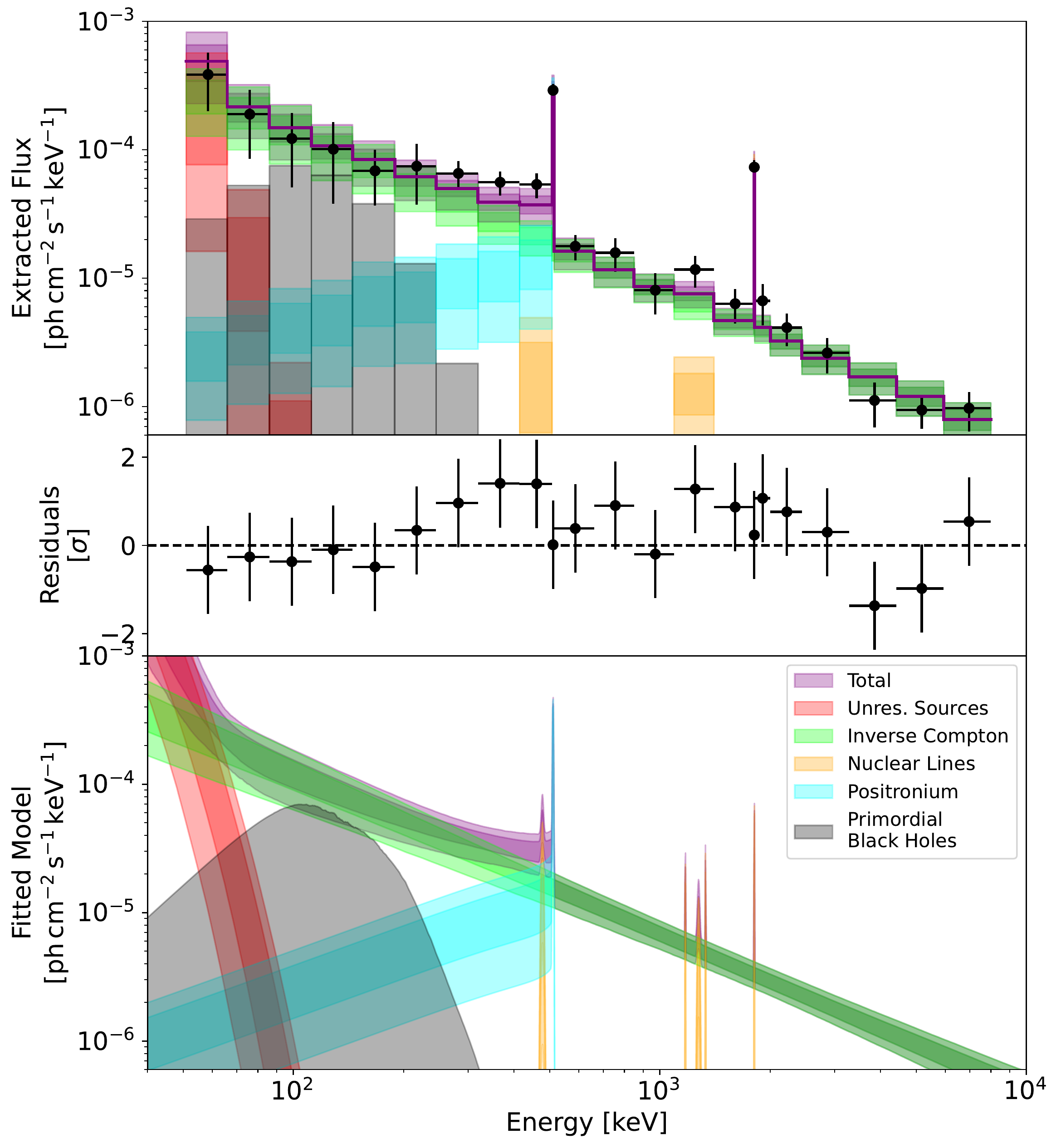}
 \caption{Extracted spectral data points and fit to the total diffuse emission spectrum (top) with residuals (center) and fitted model components (bottom). Shown are the $1$- and $2\sigma$-bands for the detected components. We also display the $2\sigma$ upper limit from our highest excluded PBH mass at $\sim 4 \times 10^{17}\,$g. See legend and main text for details.
{The tabulated spectra of the different components can be found in the public \href{https://zenodo.org/record/6505275}{Zenodo} repository.}
  }
  \label{fig:totalspectrum_fit}
\end{figure}
For the most prominent IC emision, we consider four different models and use, as input for the \spi~analysis, soft $\gamma$-ray maps obtained
from the \texttt{Galprop v56} cosmic-ray propagation code {(see~\cite{Porter:2017vaa})}.
These very same models have been used for the dedicated high-energy \spi~analysis of Ref.~\cite{Siegert:2022jii}. {The basic details
of the adopted IC models
are summarized in Appendix~\ref{sm:sys_ic}, and we also 
provide the corresponding {\it galdef} input files 
in a public \href{https://zenodo.org/record/6505275}{Zenodo} repository.}
{
We stress that the choice of these
four models, while not representing an exaustive
scan over the systematic uncertainties
induced by cosmic-ray propagation, 
is nonetheless sufficient to assess the impact
of variations of the IC morphology 
on the PBH signal evidence and bounds.
As shown in Ref.~\cite{Siegert:2022jii}, these 
models do indeed predict different IC morphologies  {(with variations also reaching up 50-60\%)}
which induce a systematic uncertainty of 30\%
at most on the extracted MeV fluxes in the range where IC is dominant.
We anticipate that the propagated effect of IC systematic uncertainty on PBH bounds is at the tens of percent level, and so, while important for the {\it interpretation} of any mismatch between \texttt{GALPROP}-based predictions and data, a more systematic scan of the cosmic-ray
production and propagation parameter space
is beyond the scope of this paper. 
}

Being interested in PBH as possible DM candidates, we assume that
their spatial number density in the Galaxy follows a typical
NFW DM profile~\cite{Navarro:1996gj}.
We use the following parameter values: 
$r_s = 9.98$ kpc for the scale radius, and 
$\rho_s = 2.2 \times 10^{-24} \, \rm g/cm^3$, 
in agreement with a recent fit to Milky Way dynamical data~\cite{Karukes:2019jwa}.

% total flux 
The diffuse $\gamma$-ray flux from Galactic PBH evaporation writes as:
% equation
\begin{equation}
    \frac{\d \Phi_\gamma}{\d E}(l, b) = \frac{f_{\rm PBH}}{4 \pi M_{\rm PBH}} \frac{\d^2 N_\gamma}{\d E \d t} \int_{\rm l.o.s.} \d s \, \rho(r(s, l, b)) \, .
    \label{eq:tot_flux}
\end{equation} 
It receives contribution
from (a) a {\it normalization} factor depending on the PBH mass and fraction of PBHs which 
may be the DM in the universe ($f_{\rm PBH}$) -- the parameter we want ultimately to constrain --,
(b) a $M_{\rm PBH}$-dependent {\it spectral} term, which defines the energy dependence of the signal, discussed below, and (c) a {\it spatial} term, which corresponds to the integral along the line of sight (l.o.s.) of the DM profile -- also known in the DM literature as $\mathcal{D}$-factor --, which determines the morphology of the signal. Note that, because of the detection technique, the analysis is insensitive to isotropic contributions, such as the extragalactic contribution or the isotropic residual of the Galactic halo, which are thus eliminated from our NFW model template.

For the sake of our analysis, we notice that the spectral and spatial terms
in Eq.~(\ref{eq:tot_flux}) factorize, so that only the l.o.s.~dependent $\mathcal{D}$-factor
enters in the definition of the PBH template used as input for the \spi~analysis (see below).
The spectral term, instead, will be fully exploited only when setting 
constraints on the PBH parameter space.

We perform the integral along any given direction in the analysis ROI
to get a map of the expected spatial distribution of the PBH signal.
The map is binned in 0.5$^\circ\times 0.5^\circ$ pixels and centered on the Galactic center for both $b$ and $l$, thus including ${\cal O}(40000)$ pixels.

%%%%%%%%%%%%%%%%%%%%%%%%%%%%%%%%%%%%%%
\section{Spectral fits of the total spectrum} 
\label{sec:spfits}
%%%%%%%%%%%%%%%%%%%%%%%%%%%%%%%%%%%%%%
{Once the components
 have been separated via angular templates, Sec.~\ref{sec:templatefit}, a spectral 
fit to the total spectrum is performed to 
derive the parameters of interest for the spectral model.}
In the spectral analysis, the astrophysical components are fitted with the following: 1)  The  population of unresolved point sources, believed to be mostly cataclysmic variables and stars with hot coronae~\cite{Lutovinov:2020ust}, is parameterized via a cutoff power-law with free normalization and cutoff energy; 2) positron annihilation, allowing for a free normalization of the 511 keV line and a free fraction of ortho-Ps controlling the continuum emission below 511 keV; 3) nuclear lines, with a free normalization; 4) IC with a power-law, with free amplitude and index.
We refer the interested reader to Appendix~\ref{sm:spectralfit} for more technical details of the spectral fit.

The PBH emission is predicted to come from {\it Hawking radiation}~\cite{10.1093/mnras/152.1.75},
with a spectrum following an almost blackbody distribution, with temperature given in natural units by $T_{\rm PBH}=M_P^2/(8\pi M_{\rm PBH})$, where $M_P$ is the Planck mass:
\begin{equation}
    \frac{\d^{2}N_{i}}{\d E \d t}=\frac{1}{2\pi}\frac{\Gamma_{i}(E,M_{\rm PBH})}{e^{E/T_{\rm PBH}}-(-1)^{2s}}\,.
    \label{eq:bh_spectrum}
\end{equation}
In the spectrum given by Eq.~(\ref{eq:bh_spectrum}), $s$ is the spin of the $i^{th}$ radiated particle, $E$ its energy and $\Gamma_{i}(E,M_{\rm PBH})$ is a species-dependent greybody factor.
To compute the spectrum of photons from Hawking evaporation of PBHs, we use \texttt{BlackHawk v1.2}~\cite{Arbey:2019mbc}. We only consider the primary spectrum since the extrapolation tables used to compute secondary photons lead to unphysical spectra in the relevant range of energies. We note that the authors of Ref.~\cite{Arbey:2019mbc} released an upgraded version (\texttt{v2.1}) of the software which includes a new tool to compute the secondary photon spectra more reliably. However, this contribution only has an effect on the low energy tail of spectra for the lowest PBH masses considered here. Therefore, our results are not significantly altered by such update of the code, as also shown in~\cite{Chen:2021ngo}.

{Spectral parameters are sampled following the \texttt{emcee} MCMC sampling method~\cite{Foreman-Mackey:2012any} within the \textit{3ML} package \cite{Vianello2015_3ML}, and allowed to vary within a broad prior range, only preventing unphysical, e.g. negative, fluxes.}
%%%%%%%%%%%%%%%%%%%%%%%%%%%%%%%%%%%%%%
\section{Results} \label{results}
\begin{figure*}[!ht]
\centering
\includegraphics[width=0.49\textwidth,trim=0.0in 0.0in 0.0in 0.0in, clip=true]{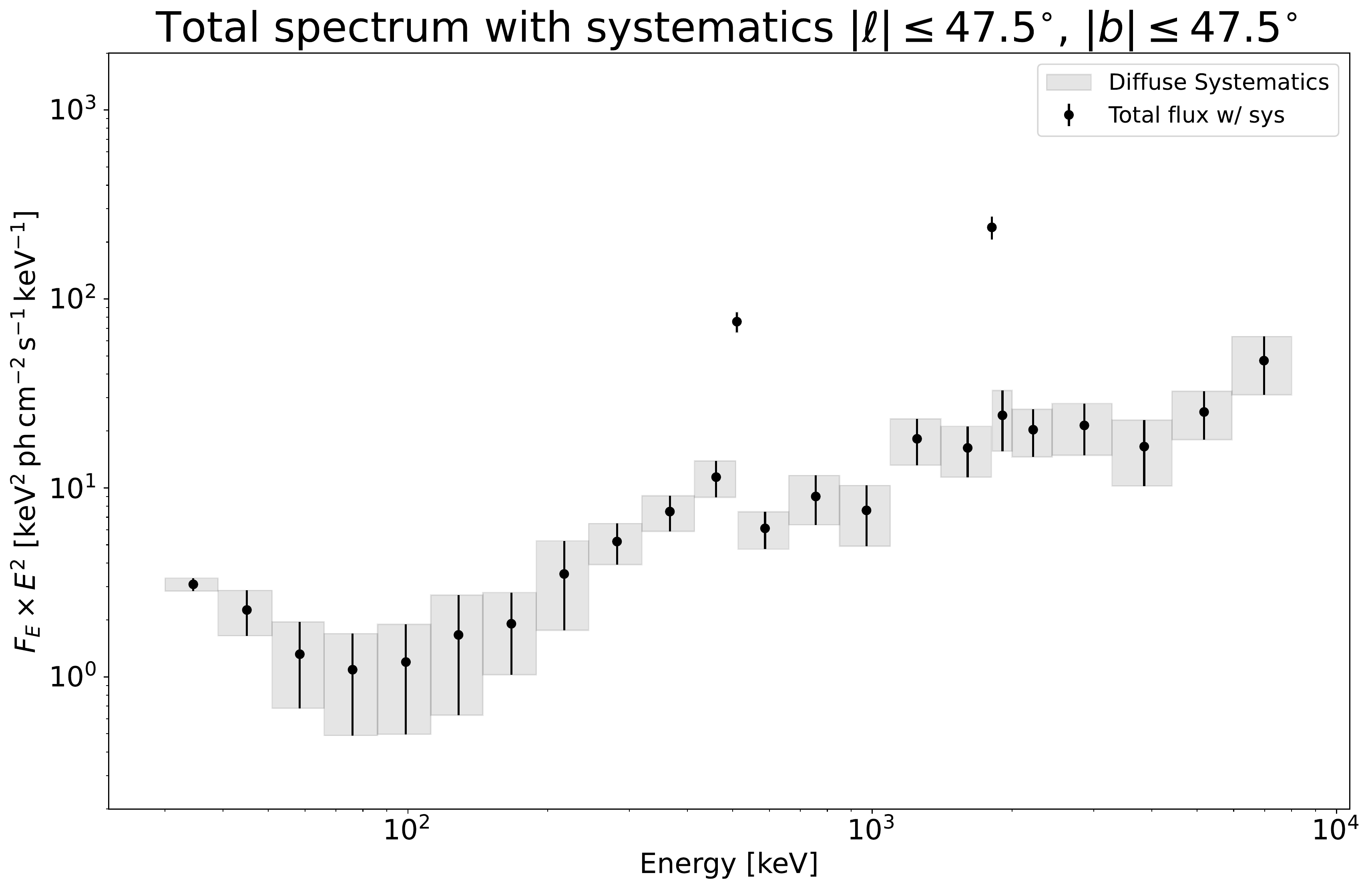}	
\includegraphics[width=0.49\textwidth,trim=0.0in 0.0in 0.0in 0.0in, clip=true]{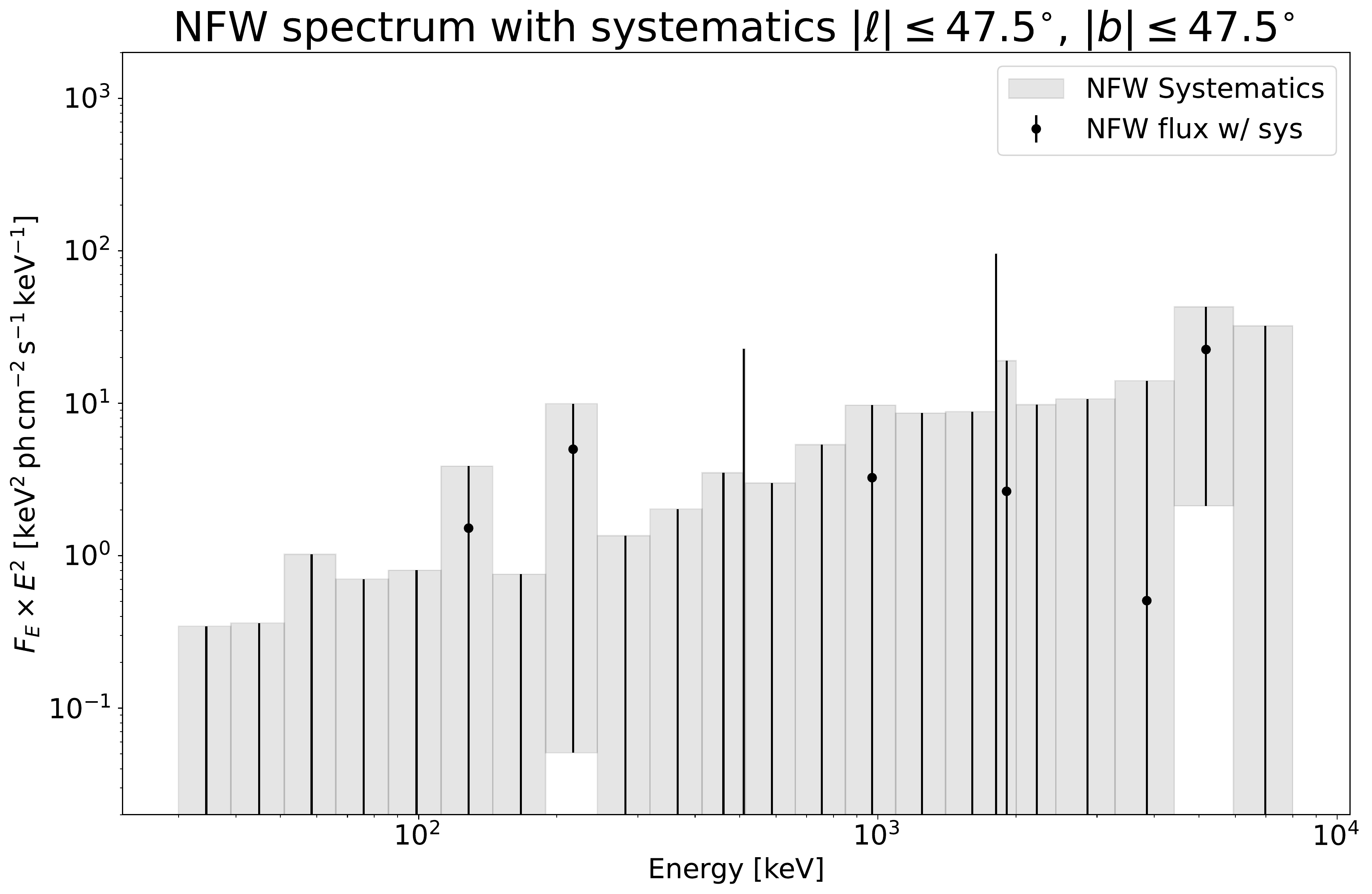}
\caption{Total (left) and NFW (right) extracted spectra including systematic uncertainties. Similar plots for the other template fit components are provides in Fig.~\ref{fig:extracted_spectra_app} in Appendix~\ref{sm:sys_ic}.}
\label{fig:extracted_spectra}
\end{figure*}

%%%%%%%%%%%%%%%%%%%%%%%%%%%%%%%%%%%%%%
{We first report our results about the angular decomposition
and template analysis to extract the flux components, c.f.~\ref{sec:templatefit}.
We first consider template components 1) - 4) in 
Sec.~\ref{sec:templatefit}, and perform the spectral extraction of the components.
}
The IC emission is the dominant contribution to the soft $\gamma$-ray diffuse emission.
All IC models perform equally well in describing \spi~data, however the 
extracted spectra show variations on the order of 5--20\,\% above 0.5\,MeV, and about 30--50\,\% below 0.5\,MeV.
We use these variations among IC models to assess
the systematic uncertainties affecting the spectral extraction.
The total flux is hardly dependent on the 
chosen IC morphology, except for the range around 100--250\,keV 
because the maximum number of templates contributes there.
We also note that in the lowest two energy bins, 30--50\,keV, the degeneracy among the fitted components is large compared to the remaining part of the spectrum, resulting in two orders of magnitude higher systematics.
Hence, we conservatively omit this band from our spectral fits.
{We show the extracted flux for the total spectrum in Fig.~\ref{fig:extracted_spectra} (left panel), including
systematic uncertainties that come from the variation of IC.
As a second step, in addition to the already known components of the Galactic flux as discussed above, we include the l.o.s.-integrated NFW profile as a template for PBH DM in the \spi~analysis, and extract the corresponding total spectrum of the diffuse soft $\gamma$-ray emission.
Fitting an NFW profile at each energy bin leads to {\it no detection} -- regardless of what IC model we adopt. 
We can therefore set 95\% C.L.~upper limits on the $\gamma$-ray flux from DM PBH, which is shown in 
Fig.~\ref{fig:extracted_spectra} (right panel).
The upper limits on the flux originating from a NFW template are robust (typically within 5--20\%, depending on energy, hence PBH mass) with respect to the chosen IC spectral model.
}

% generic model fit
{We now discuss spectral fits, c.f.~Sec.~\ref{sec:spfits}. }
Let us focus first on the total spectrum extracted including the astrophysical {
spectral} components 1) -- 4), {c.f.~Sec.~\ref{sec:spfits}}.
The fitted spectrum compares well to previous analyses, e.g. Refs.~\cite{Strong2005_gammaconti,Strong2010_CR_luminosity,Bouchet2011_diffuseCR}. 
We notice an overall good agreement, with departures from the fit within the 2$\sigma$ level, keeping in mind that some contribution expected from positron annihilation in flight, see e.g.~\cite{Beacom2006_511,Churazov2011_511,Siegert2021_RetII}, is neglected in this analysis.
We have a satisfactory fit over the full energy range, although low-energy and high-energy data seem to prefer different IC models as best fit, which may indicate a (statistically not significant) departure from a single power-law parameterisation of the IC spectrum in this range.
For a more thorough comparison in the high-energy range, we refer the 
reader to Ref.~\cite{Siegert:2022jii}, while we discuss the low-energy range in Appendix~\ref{sm:sys_ic}.

Analogous to what has been done before, we fit the total spectrum obtained including {\it all} template components with the generic model presented earlier together with the expected PBH spectrum, Eq.~(\ref{eq:tot_flux}), 
{see Fig.~\ref{fig:totalspectrum_fit}.}
As for PBH, the free spectral parameters are the PBH DM abundance, $f_{\rm PBH}$, sampled from a log-uniform distribution, and the logarithm of the PBH mass, sampled from a uniform distribution. \footnote{{Please note that this two-parametric bound is not what is done typically in the literature, i.e.~a one-parametric bound on $f_{\rm PBH}$ at a fixed mass. Our approach may lead therefore to slightly more conservative bounds.}}
Given the null evidence for the PBH component,
we achieve a comparably satisfactory fit when including PBH
with respect to the model with astrophysical components only.
This spectral fit also allows us to derive constraints on the PBH DM abundance: 
For each mass value, the constraints on the PBH DM abundance are defined such as they separate 95\% of the samples with the lowest $f_{\rm PBH}$ values from the 5\% with the highest ones.
In Fig.~\ref{fig:totalspectrum_fit}, we display the fit to the total spectrum for the highest excluded PBH mass, $M_{\rm PBH}\simeq 4 \times 10^{17}$ g.

\begin{figure}[t]
  \includegraphics[width=0.49\textwidth]{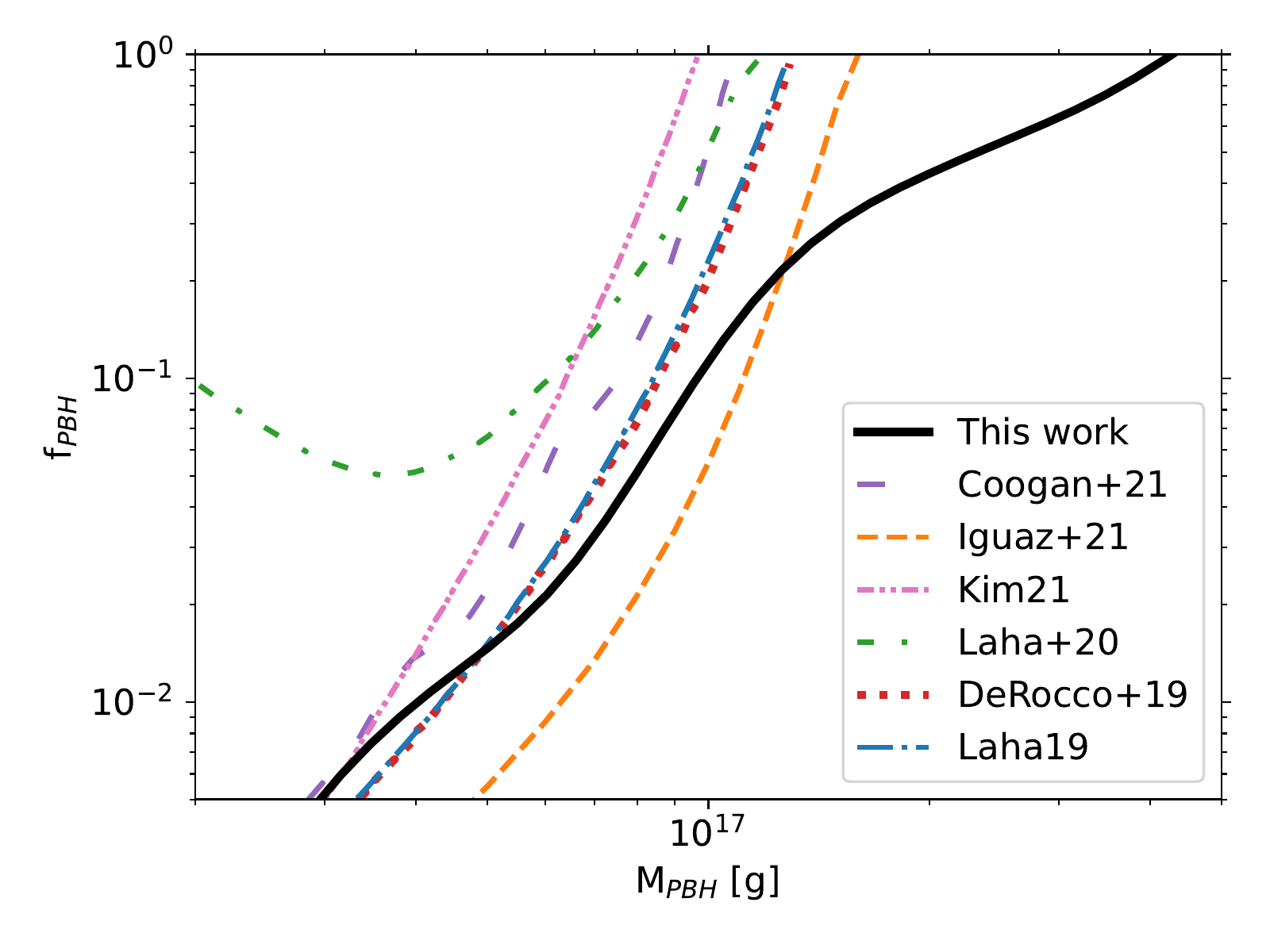}
  \caption{95\% C.L.~upper limits on the fraction $f_{\rm PBH}$ of DM that can be composed of PBHs as a function of the PBH mass, $M_{\rm PBH}$, derived in this work
   {(black solid). We also show, from top to bottom, bounds from the literature derived in: \cite{Coogan:2020tuf} (purple), \cite{Iguaz:2021irx} from the cosmic X-ray background (orange), \cite{Kim:2020ngi} (pink), \cite{Laha:2020ivk} from diffuse soft $\gamma$ rays using published \spi~results (green), \cite{DeRocco:2019fjq} (red) and \cite{Laha:2019ssq} from 511 keV constraint (NFW ``3 kpc'' case, blue).}
}
  \label{fig:PBHlimits}
\end{figure}

In Fig.~\ref{fig:PBHlimits}, we show our main result: We exclude
that PBHs account for all the DM in the universe up to masses of $\simeq 4 \times 10^{17}$ g. 
Compared to existing bounds,  {a selection of which is} of which are also reported in Fig.~\ref{fig:PBHlimits}~\footnote{{Further bounds relevant in this mass-range are in~\cite{Boudaud:2018hqb,DeRocco:2019fjq,Coogan:2020tuf}; other bounds from medium {\it heating} only extend to lower masses, but can go deeper in $f_{\rm PBH}$ space, see~\cite{Kim:2020ngi,Laha:2020vhg}.}}, ours excludes the largest  PBH masses in the low-mass window to date,
and significantly closes in to the right into the asteroid mass range. Note that the comparatively weaker bounds between $2\times 10^{16}\,$ g and $ 10^{17}\,$g are due to the fact that for most masses in this range a PBH contribution would partially fill-in the small excesses in the residuals between 300 keV and a few MeV visible in Fig.~\ref{fig:totalspectrum_fit}.

%%%%%%%%%%%%%%%%%%%%%%%%%%%%%%%%%%%%%%
\section{Discussion and conclusions} \label{discconcl}
%%%%%%%%%%%%%%%%%%%%%%%%%%%%%%%%%%%%%%
The results obtained are robust against a number of systematic checks we performed. For instance, removing the first two energy bins (which are affected by the largest systematic uncertainties) from the total spectrum does not affect the limits. 
The limits are also robust against the choice of the prior on the PBH mass down to $1.8 \times 10^{16}$ g.
Furthermore, the limits are not very sensitive to the specific model of IC, among the ones we tested. We stress, however, that even in the case in which we neglect any astrophysical background spectral information in the limit-setting procedure {and use the extracted NFW flux presented in Fig.~\ref{fig:extracted_spectra}}, the limit would still exclude $f_{\rm PBH}=1$ at $M_{\rm PBH} = 2\times 10^{17}$ g, i.e. would still remain the strongest limit on PBH constituting the totality of DM. 

As common to all Galactic limits in the literature, translating the bounds on flux into bounds on $f_{\rm PBH}$ carries on the standard uncertainty related to the local DM density, of about 30\%. Since we are focusing on a very large ROI (as opposed to the innermost Galaxy), the uncertainty due to the mere shape of the DM halo is mild: By considering a cuspier halo model (resp. cored model) according to Tab.~1 of Ref~\cite{Calore:2022stf}, we would infer 16\% tighter (resp. 10\% looser) bounds.

We have not considered here 
photons that come from positrons, produced by PBH evaporation, annihilating  
with electrons of the interstellar medium 
in flight or at rest, and 
generating an additional diffuse continuum contribution or 511\,keV line signal, respectively, as was performed in Ref.\,\cite{Siegert2021_RetII} for the case of the dwarf galaxy Reticulum II. 
Also, we neglect diffuse photons that come 
from positrons and electrons from PBHs which 
induce IC emission when scattered off by 
low ambient photons of the interstellar radiation field. 
We indeed explicitly checked that this
contribution does not contain any additional constraining power for the \spi~analysis:
The MeV spectrum is sensitive only to the overall number of electrons and positrons injected and the level of re-acceleration in the propagation model, and it is therefore highly uncertain as a probe of PBHs.

Also, we presented bounds for a monochromatic mass function. Relaxing this condition is expected to lead to mildly more stringent bounds. For instance, we checked that a lognormal mass distribution centered at $M_{\rm PBH} = 10^{17}\,$g with a dimensionless width $\sigma\simeq 0.1$ would lead to $\sim 10\%$ tighter bounds, which improve to a factor of $\sim$ 2 stronger bounds when $\sigma\gtrsim 0.3$. More stringent bounds are also obtained if going from non-rotating (Schwarzschild) PBH to rotating (Kerr) PBH, see e.g.~\cite{Iguaz:2021irx}.

Let us briefly comment on some directions for improvements:
On the one hand, the new SPI data points can be put into context of {\it Fermi}-LAT{, COMPTEL~\cite{Grenier:2015egx}} and eROSITA~{\cite{eROSITA:2012lfj}} diffuse emission measurements so to evaluate how the IC modeling can be improved, and, ultimately, what are the consequences for PBH DM. {In particular, if the diffuse emission in the keV band can be separated from the point sources, eROSITA might probe PBHs with masses around $\sim 10^{18}\,$g, beyond current reach}. On the other hand, if additional components, such as positron annihilation at rest or in flight from PBH are included, the limits are expect to be stronger. Nonetheless, computing the expected 511 keV signal is challenging given the highly uncertain propagation conditions in the interstellar medium at those energies.

Finally, we highlight that the new \spi~template analysis developed here, which includes 
a template component for NFW DM, can be straightforwardly extended to set constraints on 
other generic models for decaying particle DM and it is thus of broad interest. {This strategy can also prove useful in pushing the sensitivity expected by future missions, as studied e.g. in~\cite{Coogan:2020tuf,Ray:2021mxu,Auffinger:2022dic}.}

\medskip
%%%%%%%%%%%%%%%%%%%%%%%%%%%%%%%%%%%%%%
\textbf{Acknowledgments.} 
F.C., J.B.~and P.D.S. acknowledge support by the ``Agence Nationale de la Recherche'', grant n. ANR-19-CE31-0005-01 (PI: F. Calore). T.S. acknowledges support by the Bundesministerium f\"ur Wirtschaft und Energie via the Deutsches Zentrum f\"ur Luft- und Raumfahrt (DLR) under contract number 50 OX 2201.
This work made use of the \texttt{Galprop} code for cosmic-ray propagation, \url{https://galprop.stanford.edu/}.
%%%%%%%%%%%%%%%%%%%%%%%%%%%%%%%%%%%%%%

\appendix
%%%%%%%%%%%%%%%%%%%%%%%%%%%%%%%
\section{INTEGRAL/SPI data set}
\label{sm:dataset}
%%%%%%%%%%%%%%%%%%%%%%%%%%%%%%%
% Any relevant information about the data set used,cleaning, etc
The SPI data set used here is identical to the one from Ref.\,\cite{Siegert:2022jii} with an extension to lower energies, thus in total from 30\,keV to 8\,MeV.
For completeness, we summarize the main characteristics in the following:
We define 22 logarithmic energy bins and include two narrow bins to account for the 511\,keV line from positron annihilation and the 1809\,keV line from $^{26}${Al} decay.
The two $^{60}${Fe} lines at 1173 and 1332\,keV only show a significance of $\sim 5\sigma$ above the IC continuum when analyzed in combination \cite{Wang2020_Fe60}.
We therefore omit cutting the intermediate energy bin from 1093 to 1404\,keV into five individual bins a this would unnecessarily increase the uncertainties.
Instead, we treat the $^{60}${Fe} lines in relation to the $^{26}${Al} line (see Sec.\,\ref{sm:spectralfit}) because they are expected to contribute to the flux in this energy bin.
Other nuclear lines that might contribute to the diffuse emission such as from $^{7}${Be} at 478\,keV and $^{22}${Na} at 1275\,keV are also absorbed in the broad logarithmic energy bins, and treated individually in the spectral fit because their contributions were found to be negligible \cite{Siegert2021_BHMnovae}.

The total number of targeted observations in this data set is 35892 pointings for the range above 514\,keV and 34428 pointings for the range below.
The data set includes a dead-time corrected total exposure time of 65.3--68.5\,Ms, depending on the energy range chosen, within a spherical rectangle of $\Delta l \times \Delta b = 95^{\circ} \times 95^{\circ}$ centered at the Galactic center.
{To account for failures of the
Germanium detector, we separated the data set in five different epochs, one for each camera configuration. The relative normalizations of the five different imaging responses are fixed by the official IRF distribution from the ISDC.}

The background variability time scale from the background handling method in Ref.\,\cite{Siegert2019_SPIBG} changes quickly towards a higher variability the smaller the energy.
We provide an overview of the number of fitted parameters, included components, degrees of freedom, background variability, and fit quality in Tab.\,\ref{tab:dataset}.
Except for the the energy bin 39--51\,keV with the highest background variability and largest number of point sources, the all energy bins show an adequate fit quality as measured by the reduced $\chi^2$.
 {Extracting the flux values and their uncertainties as a function of energy occurs in the native SPI data space, i.e. number of photons recorded per detector, energy, and pointing.
Spectral fits (see Appendix\,\ref{sm:spectralfit}) are then performed in the reconstructed spectral domain.}

\begin{table}[!t]
        \centering
        \begin{tabular}{c|rrrr|rr}
            \hline\hline
            Energy band     & $n_{\rm data}$ & $T_{\rm BG}$ & $\mathrm{dof}$ & $\chi^2/\mathrm{dof}$ & $n_{\rm PS}$ & Proc. \\
            \hline
            $39$--$51$ & $556184$ & $0.09$ & $530703$ & - & 109 & SE \\
            $51$--$66$ & $556184$ & $0.19$ & $542774$ & $1.1328$ & 92 & SE \\ 
            $66$--$86$ & $556184$ & $0.19$ & $542810$ & $1.1622$ & 56 & SE \\ 
            $86$--$112$ & $556184$ & $0.19$ & $542816$ & $1.0889$ & 50 & SE \\ 
            $112$--$145$ & $556184$ & $0.75$ & $551309$ & $1.0137$ & 44 & SE \\ 
            $145$--$189$ & $556184$ & $0.75$ & $551321$ & $1.0139$ & 32 & SE \\ 
            $189$--$245$ & $556184$ & $1.5$ & $553502$ & $1.0135$ & 21 & SE \\ 
            $245$--$319$ & $556184$ & $0.75$ & $551342$ & $1.0038$ & 11 & SE \\ 
            $319$--$414$ & $556184$ & $1.5$ & $553515$ & $1.0029$ & 8 & SE \\ 
            $414$--$508$ & $556184$ & $1.5$ & $553518$ & $1.0062$ & 5 & SE \\ 
            $508$--$514$ & $556184$ & $3$ & $554707$ & $0.9912$ & 4 & SE \\    
            \hline
            $514$--$661$ & $578764$ & $0.75$ & $573827$ & $1.0059$ & 4 & PSD \\
            $661$--$850$ & $578764$ & $0.75$ & $573827$ & $0.9984$ & 4 & PSD \\
            $850$--$1093$ & $578764$ & $0.75$ & $573831$ & $0.9974$ & - & PSD \\
            $1093$--$1404$ & $578764$ & $0.75$ & $573831$ & $0.9974$ & - & PSD \\
            $1404$--$1805$ & $578764$ & $1.5$ & $576047$ & $0.9939$ & - & PSD \\
            $1805$--$1813$ & $578764$ & $3$ & $577254$ & $0.9935$ & - & PSD \\
            $1813$--$2000$ & $578764$ & $3$ & $577255$ & $0.9953$ & - & PSD \\
            \hline
            $2000$--$2440$ & $582349$ & $6$ & $581390$ & $1.0057$ & - & HE \\
            $2440$--$3283$ & $582349$ & $3$ & $580836$ & $1.0040$ & - & HE \\
            $3283$--$4418$ & $582349$ & $3$ & $580836$ & $1.0026$ & - & HE \\
            $4418$--$5945$ & $582349$ & $3$ & $580836$ & $1.0064$ & - & HE \\
            $5945$--$8000$ & $582349$ & $6$ & $581390$ & $1.0038$ & - & HE \\
            \hline
        \end{tabular}
        \caption{Dataset characteristics. The columns from left to right are the energy band in units of keV, the number of data points, the background variability timescale in units of days, the corresponding number of dof, the calculated reduced $\chi^2$ value from the best fit, the number of used point sources, and the SPI processing chain. The values from 514 to 8000\,keV are the same as in Ref.\,\cite{Siegert:2022jii} and only repeated here for consistency.}
        \label{tab:dataset}
\end{table}

\section{Extracted spectra and systematic uncertainties}
\label{sm:sys_ic}
\begin{figure*}
    \centering
    \includegraphics[width=0.24\textwidth]{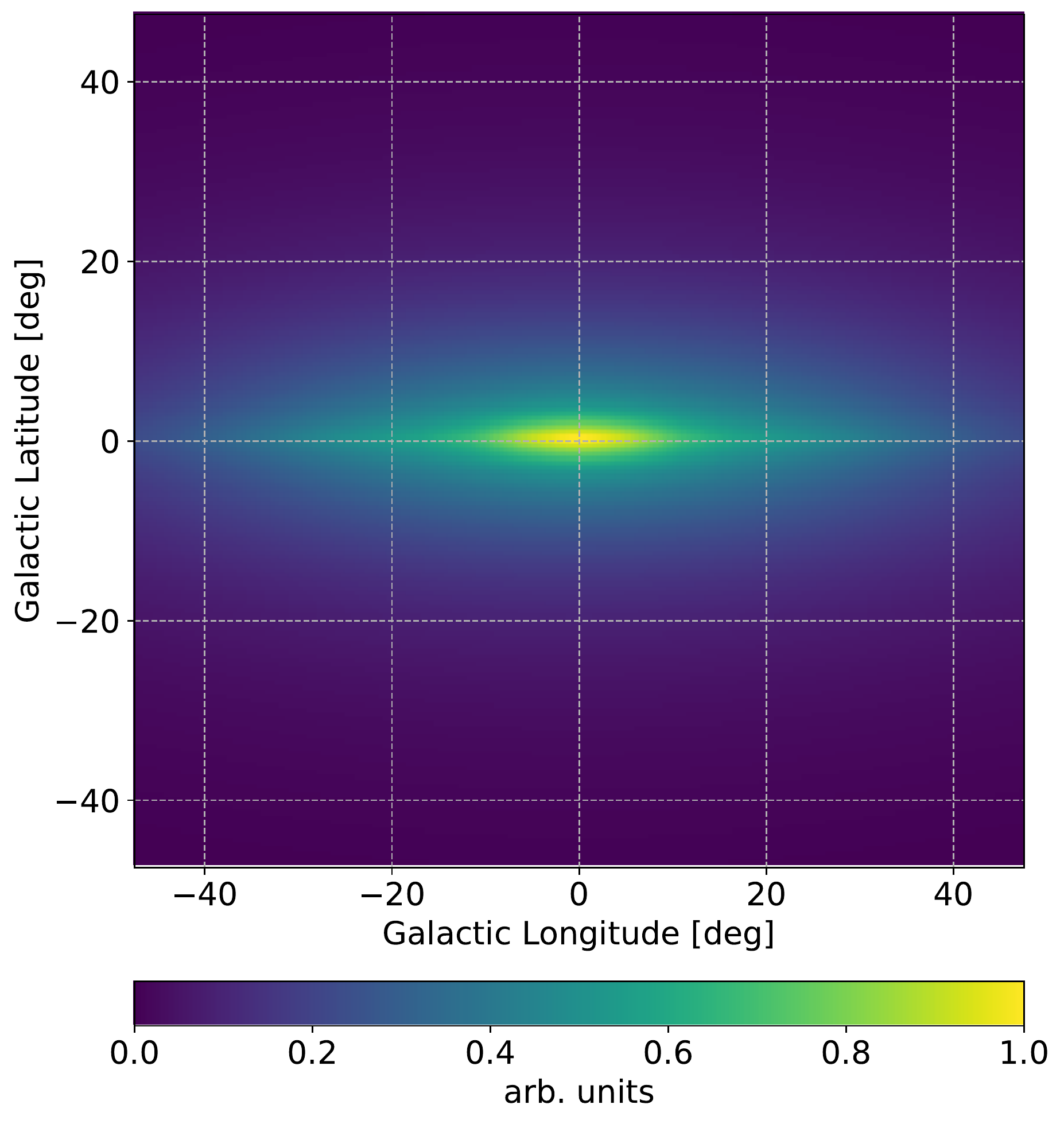}
    \includegraphics[width=0.24\textwidth]{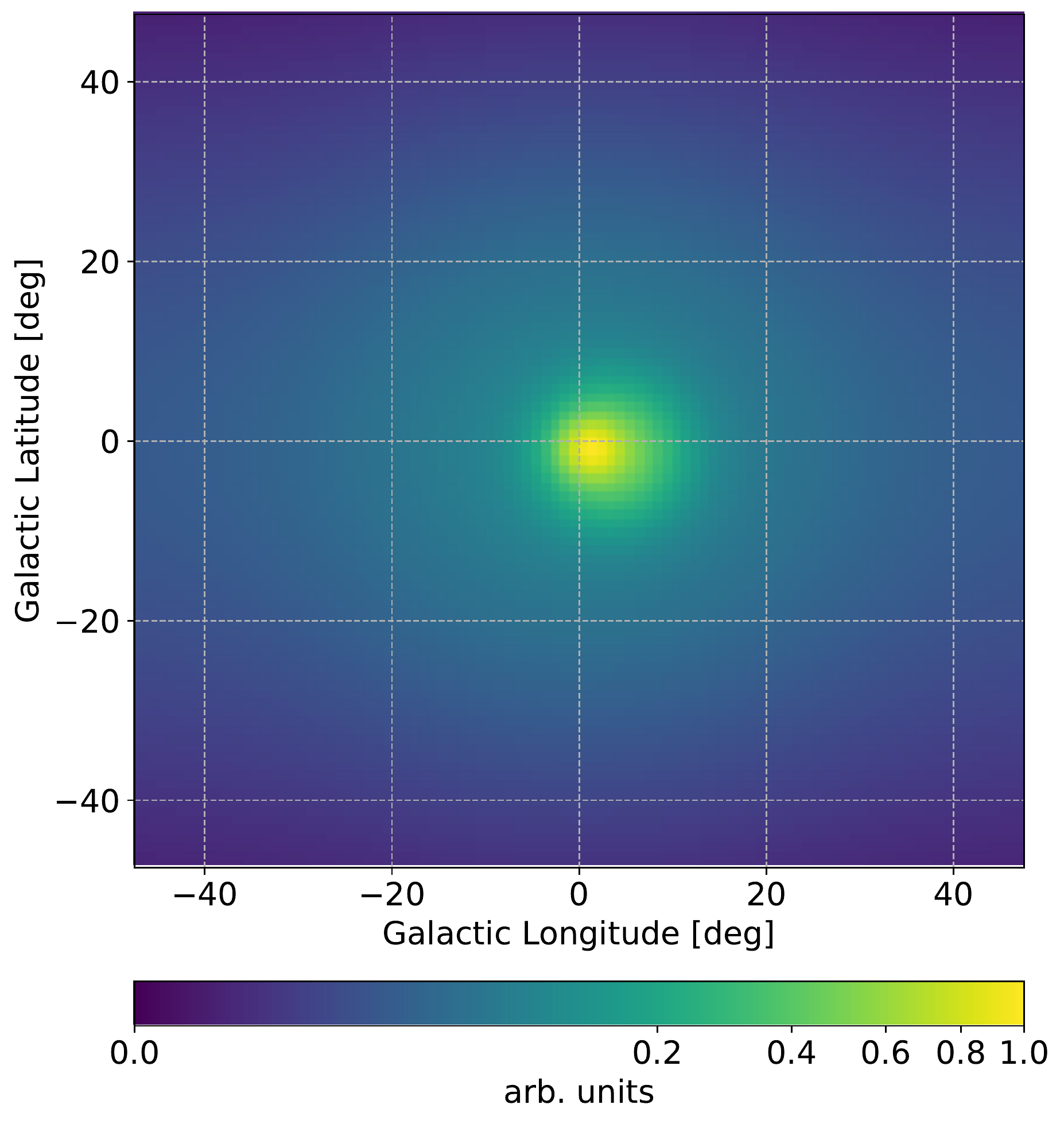}
    \includegraphics[width=0.24\textwidth]{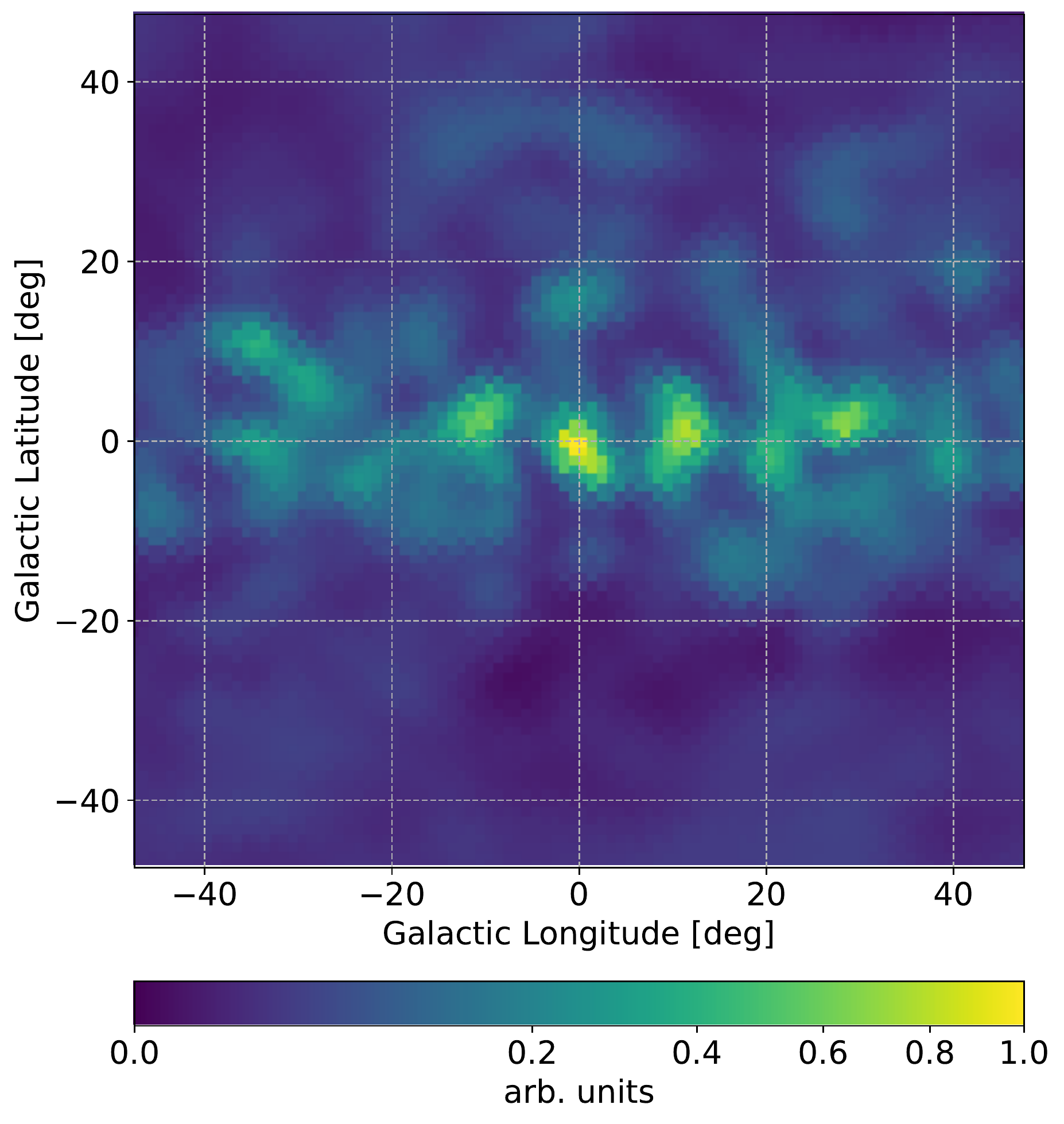}
    \includegraphics[width=0.24\textwidth]{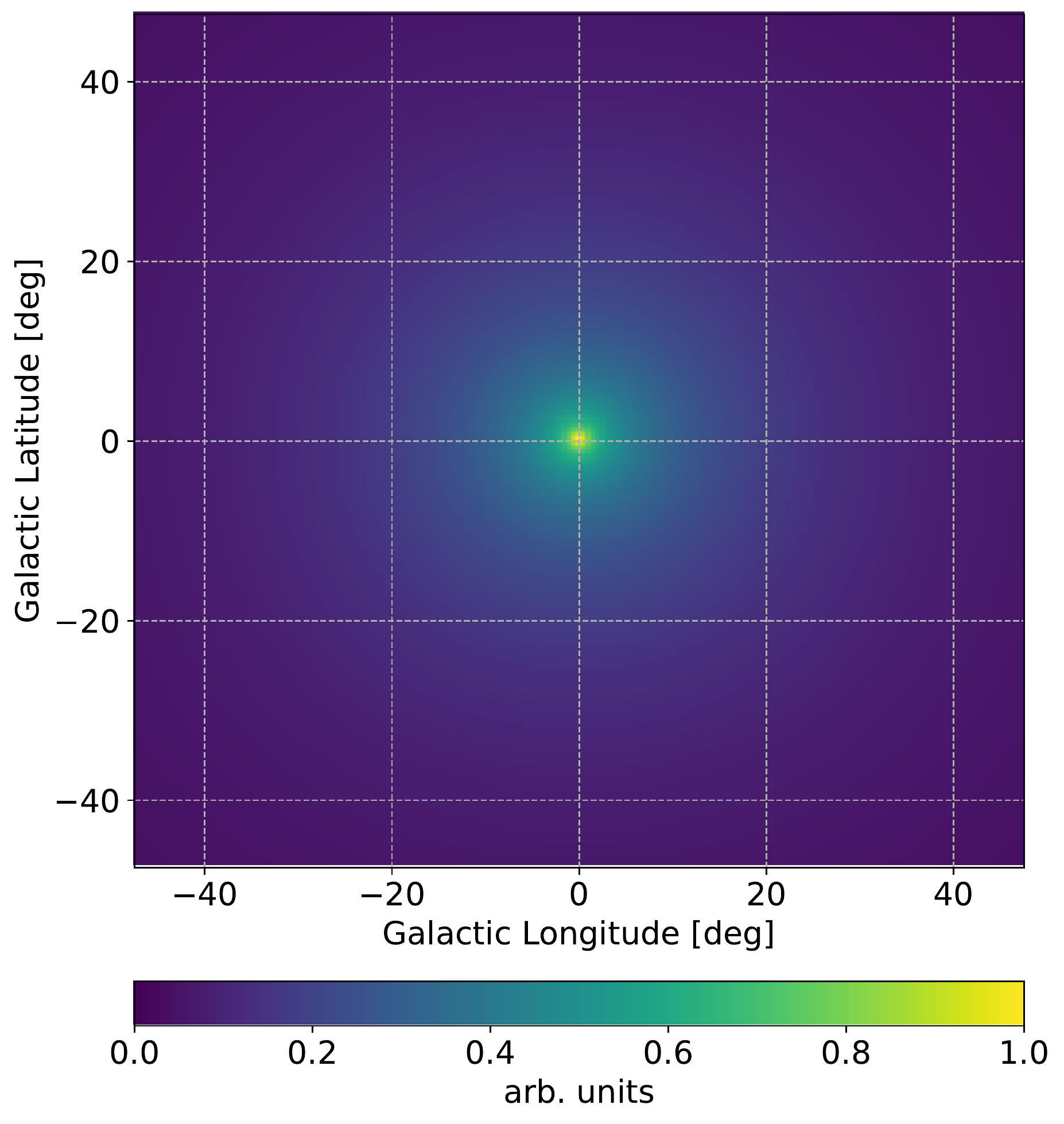}
    \caption{Sky models used for the analysis. Shown are, from left to right, the IC emission model at 281\,keV from the model variant $\delta_1 = \delta_2 = 0.5$ (see Ref.\,\cite{Siegert:2022jii}), the Positronium emission model from Ref.\,\cite{Siegert2016_511}, the 1.8\,MeV $^{26}$Al map reconstructed from SPI data by Ref.\,\cite{Bouchet2015_26Al}, as well as a line-of-sight integrated NFW profile (see main text).}
    \label{fig:diffuse_models}
\end{figure*}

\begin{figure}
    \centering
    \includegraphics[width=0.23\textwidth]{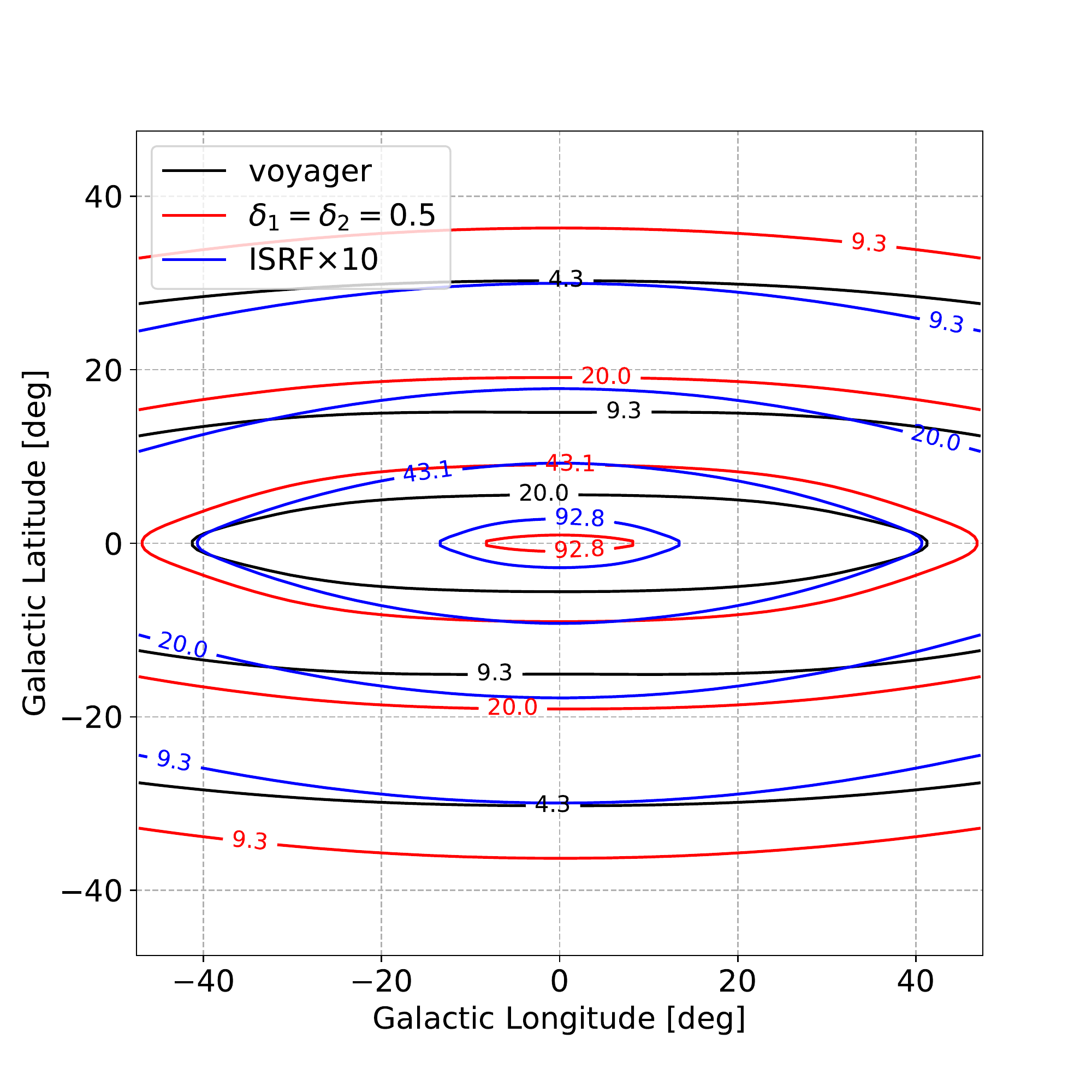}
    \includegraphics[width=0.23\textwidth]{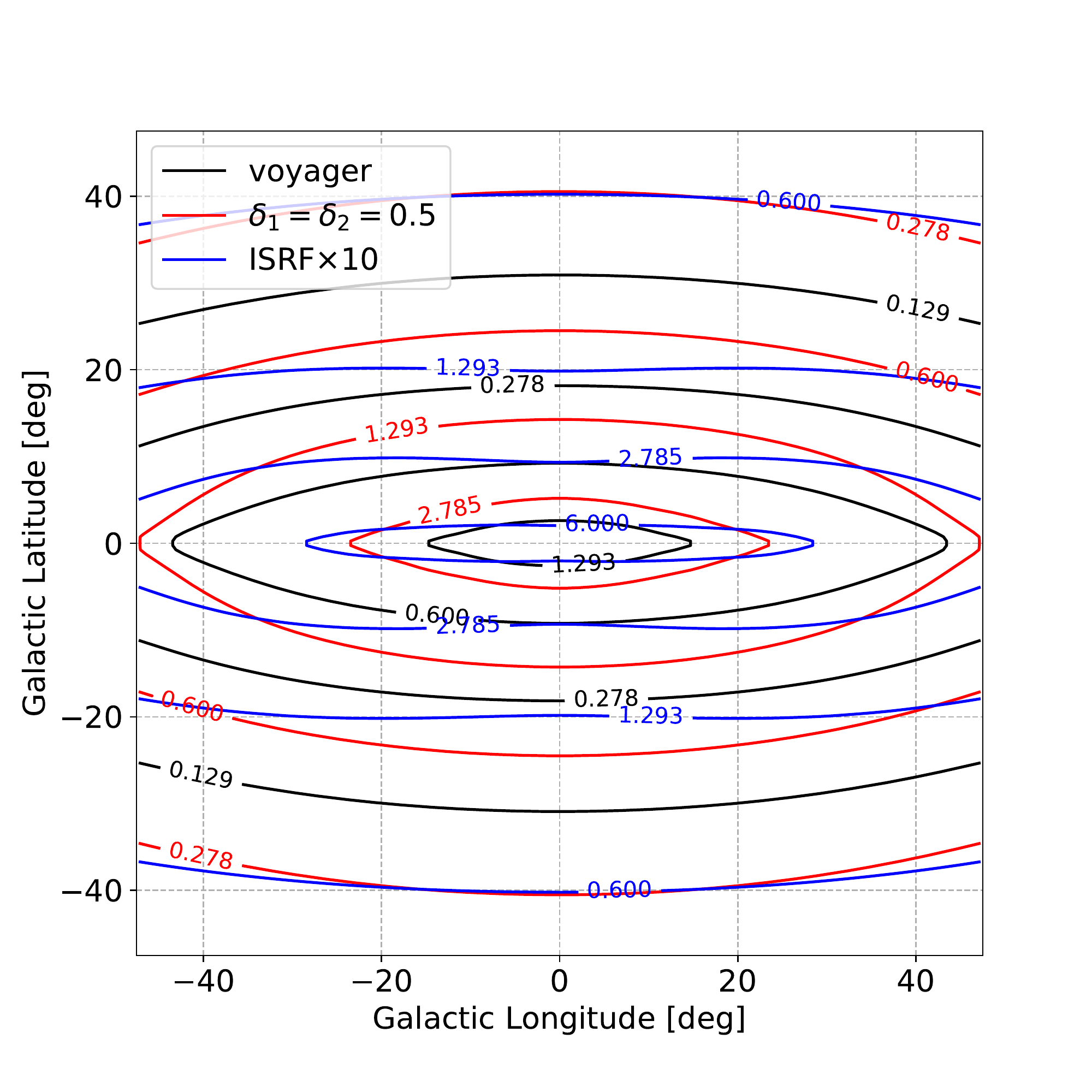}
    \caption{Comparison of three different IC model configurations for 30\,keV (left) and 8000\,keV (right). The units of the contours are $10^{-4}\,\mathrm{ph\,cm^{-2}\,s^{-1}\,sr^{-1}}$ and $10^{-7}\,\mathrm{ph\,cm^{-2}\,s^{-1}\,sr^{-1}}$ for the 30 and 8000\,keV figures, respectively.}
    \label{fig:IC_comparisons}
\end{figure}

{In Fig.\,\ref{fig:diffuse_models} we show the diffuse emission models, i.e.~templates, used in this work, and used for the spectral extraction, Sec.~\ref{sec:templatefit}.}

The systematic study of which diffuse emission model for the Inverse Compton (IC) scattering component in the Milky Way best describes the lower energies of the SPI spectrum (30--514\,keV) is detailed below.
It follows the same thread as in a recent work about SPI's high-energy range (514--8000\,keV) from Ref.\,\cite{Siegert:2022jii}.
We briefly summarize here the main features of the IC models we used.
We consider the model  {$\mathrm{^SS ^Z4 ^R20 ^T150 ^C5}$} as defined in \cite{Ackermann2012_FermiLATGeV} which provides a good fit to {\it Fermi}-LAT $\gamma$-ray data.
 {Based on the {\it Fermi} galdef file, we then create a model matching the parameters} in table 2 of \cite{Bisschoff2019_Voyager1CR}, which is tuned to match low-energy cosmic-ray data from $\mathrm{Voyager}$\,1 (\texttt{voyager} baseline).
To assess the systematic uncertaities from IC, we also define two additional models obtained varying some parameters of the \texttt{voyager} baseline: (i) $\delta_1 = \delta_2 = 0.5$, case of a single diffusion index as suggested by \cite{Genolini2019_AMS02_B2C,Weinrich2020_AMS02_halosize}; 
(ii) \texttt{ISRF$\times$10}, case with a factor 10 stronger optical ISFR).
 {All four models inherit the same cosmic-ray source distribution from $\mathrm{^SS ^Z4 ^R20 ^T150 ^C5}$, i.e.~the SNR distribution from~\cite{Case_1998}.}
{We made the GALPROP input, i.e.~galdef, files
of the models considered publicly available 
at~\href{https://zenodo.org/record/6505275}{Zenodo}.}

 {Another source of variations of the IC morphology can be, e.g., a different distribution of cosmic-ray sources. We explicitly checked that 
the variations induced on the IC morphology by assuming distributions L, Y, and O from~\cite{Acero2016_FermiLAT_sources} are at the level of 15-20\% at most with respect to our \texttt{voyager} baseline model. Therefore, they cannot dominate the systematic uncertainty of the extracted PBH flux, and, in turn, significantly alter the PBH bounds.
Additionally, other untested effects are
in-homogeneous diffusion and other propagation scenarios, e.g.~addressed in~\cite{2016MNRAS.462L..88R, 2017JCAP...10..019C,2018MNRAS.475.2724O, 2019PhRvD..99d3007O}.
We stress that, because of the poor SPI angular resolution and contrary to what occurs 
at GeV energies, different models at MeV energies carry larger degenerancies which are difficult to break, see Ref.~\cite{Siegert:2022jii} for a thorough discussion.
}

Within the systematic uncertainties, our IC model selection (\texttt{voyager} baseline, $\delta_1 = \delta_2 = 0.5$, \texttt{ISRF$\times$10}, $\mathrm{^SS ^Z4 ^R20 ^T150 ^C5}$)   
is consistent with previous studies, e.g. Ref.\,\cite{Bouchet2011_diffuseCR} within $\sim 25\,\%$.
There is no single best IC model for the entire energy range of 30--8000\,keV.
At higher energies ($> 500$\,keV) model variant $\delta_1 = \delta_2 = 0.5$ fits best, whereas at lower energies \texttt{voyager} baseline and $\mathrm{^SS ^Z4 ^R20 ^T150 ^C5}$ are closest to the extracted flux data points from SPI.
{Because IC shows a known variation with energy, and is, in addition, not well determined in the MeV range, we show, as an example, the differences in three of our models for 30 and 8000\,keV, respectively, in Fig.\,\ref{fig:IC_comparisons}.}

\begin{figure*}[!ht]
\centering
\includegraphics[width=0.49\textwidth,trim=0.0in 0.0in 0.0in 0.0in, clip=true]{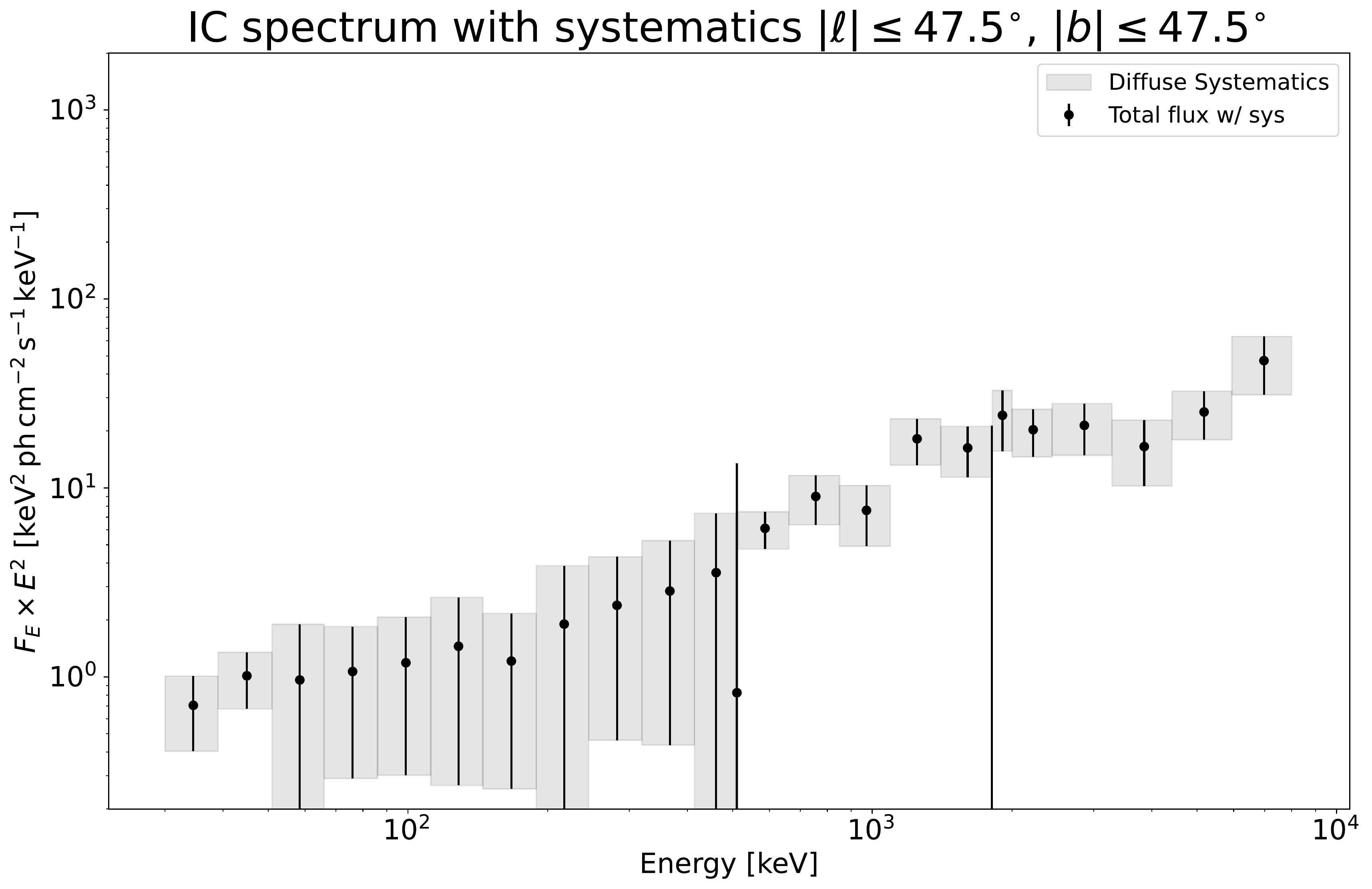}
\includegraphics[width=0.49\textwidth,trim=0.0in 0.0in 0.0in 0.0in, clip=true]{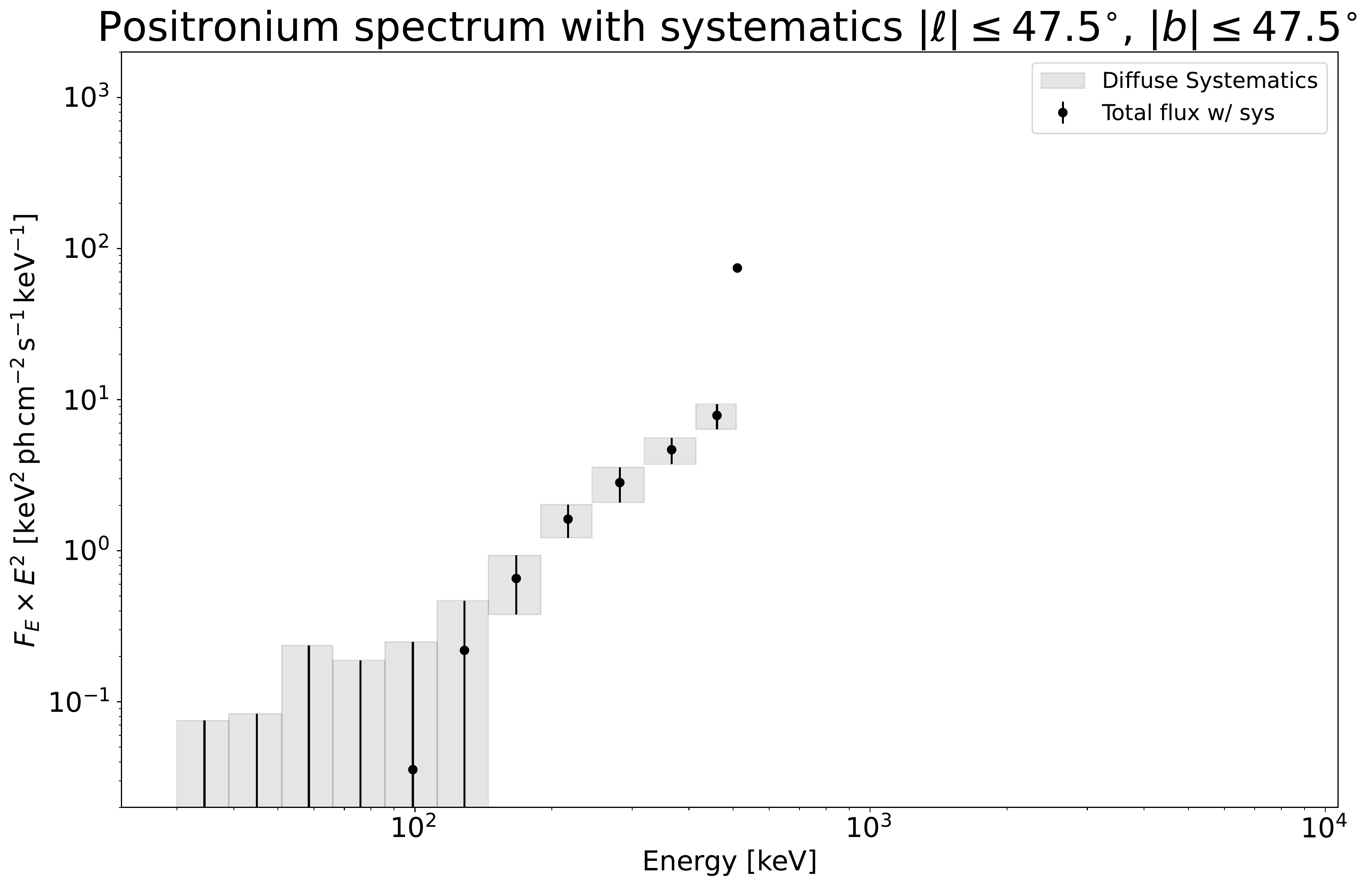}\\
\includegraphics[width=0.49\textwidth,trim=0.0in 0.0in 0.0in 0.0in, clip=true]{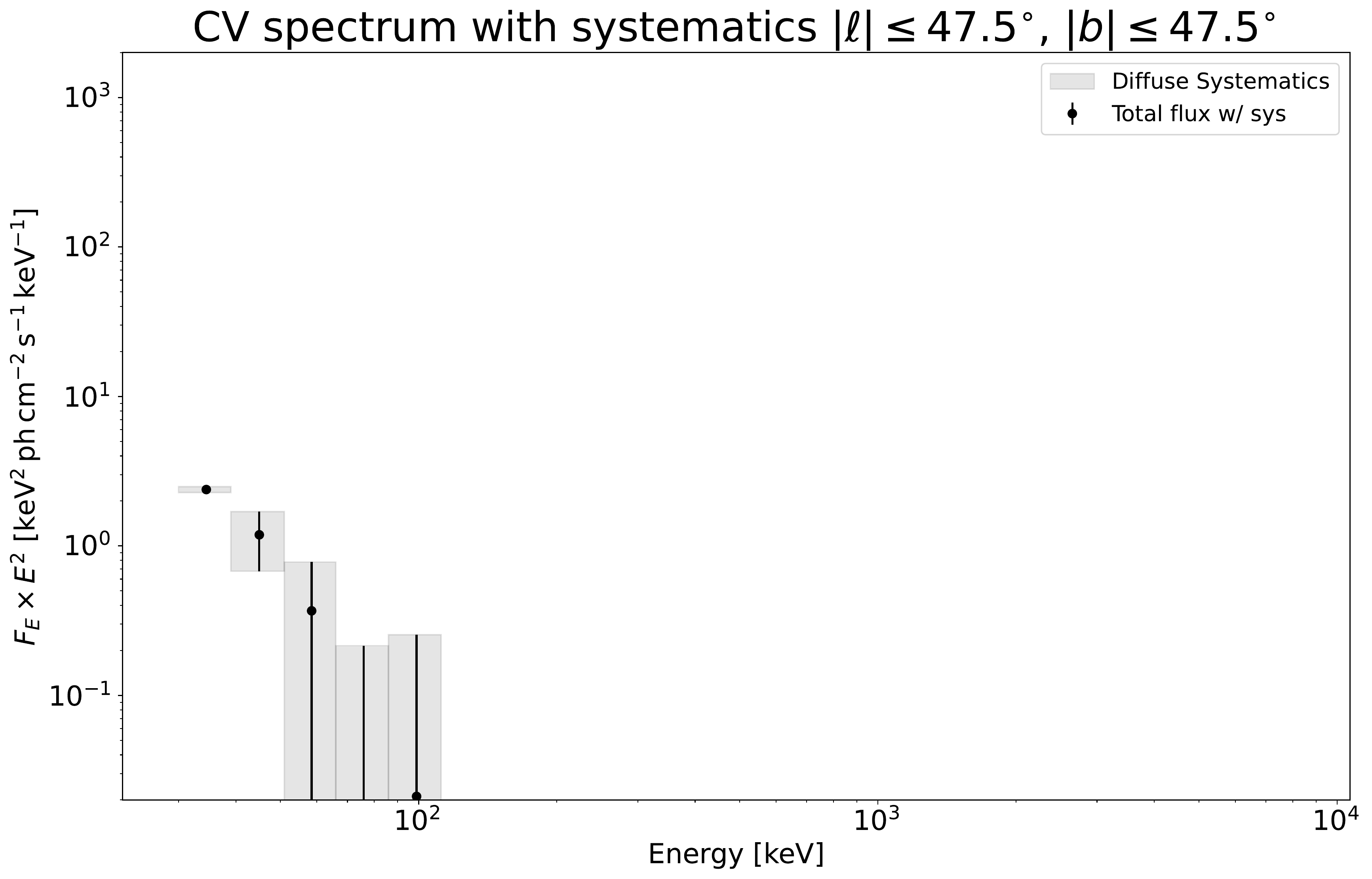}
\caption{Final spectra including systematics: top left: IC, top right: Ps, bottom: CVs.}
\label{fig:extracted_spectra_app}
\end{figure*}

In Fig.\,\ref{fig:extracted_spectra_app}, the extracted spectra with systematics are shown for the individual components {not presented in Fig.~\ref{fig:extracted_spectra}}.
{As for the NFW template,} there is no detection that could resemble the the expected shape of PBH evaporation.
A few energy bins (112--145\,keV, 189--245\,keV, and 4418--5945\,keV) show more than $3\sigma$ deviations from zero.
However, including the different IC model variants for the systematics in the NFW spectrum, none of the energy bins shows an excess larger than $1.2\sigma$.

\section{Spectral fit to extracted spectral points}
\label{sm:spectralfit}
After the spectral extraction per energy bin, the total and component-wise flux data points are analyzed by a spectral fit including the energy redistribution of SPI.
This means (astro)physical or empirical models that are parametrised through spectral parameters to predict a differential flux in units of $\mathrm{ph\,cm^{-2}\,s^{-1}\,keV^{-1}}$ are convolved with the energy redistribution matrix to account for possible dispersion in the measurement.
Thus the differential flux is converted to a rate of photons expected during the measurement time.
As an example here, we detail out the spectral model used for the total spectrum; the individual components follow accordingly.

We describe the total spectrum as a linear combination of a) a cut-off power-law to describe the population of unresolved point sources in the Galactic ridge (mainly cataclysmic variables), b) a power-law to describe the IC scattering component from GeV electrons, c) the Positronium emission from the annihilation of cooled cosmic-ray positrons with electrons in the interstellar medium, d) nuclear lines from massive star and nova ejecta, and e) the possible contribution from PBHs.
From a) to e), the models read
\begin{eqnarray}
    & \mathrm{a)~} & C_0 \left( \frac{E}{E_0} \right)^{\alpha_0} \exp\left( -\frac{E}{E_C} \right) \\
    & \mathrm{b)~} & C_1 \left( \frac{E}{E_1} \right)^{\alpha_1} \\
    & \mathrm{c)~} & f(f_{\rm Ps},F_{511}) \\
    & \mathrm{d)~} & \sum_{i} \left[ \frac{F_{i}}{\sqrt{2\pi}\sigma_i} \exp\left( -\frac{1}{2} \left( \frac{E-\mu_i}{\sigma_i} \right)^2 \right) \right] \\
    & &  \mathrm{with~} F_{60.1} = F_{60.2} = b F_{26} \nonumber \\
    & \mathrm{e)~} & g(f_{\rm PBH},M_{\rm PBH})\mathrm{,~see~main~paper.}
\label{eq:fit_functions}
\end{eqnarray}
Here, the following parameters are fixed because they can either not be constrained through the spectrum alone or are mere pivotal parameters:
$E_0 = 50$\,keV, $E_1 = 1000$\,keV, $\alpha_0 = 0$, $\mu_{26} = 1809$\,keV, $\mu_{60.1} = 1173$\,keV, $\mu_{60.2} = 1332$\,keV, $\mu_{7} = 478$\,keV, $\mu_{22} = 1275$\,keV, $\sigma_{26} = 1.7$\,keV, $\sigma_{60.1} = 1.5$\,keV, $\sigma_{60.2} = 1.5$\,keV, $\sigma_{7} = 2.4$\,keV, and $\sigma_{22} = 8.5$\,keV.
 {For all parameters we use log-uniform priors in a broad range, except for the scaling between the ${}^{26}$Al and the ${}^{60}$Fe lines with $b = 0.18 \pm 0.08$ (see \cite{Wang2020_Fe60}), and the nova lines from ${}^{7}$Be and ${}^{22}$Na for which we use} 
 a truncated Gaussian prior bound to zero and width of $2.0 \times 10^{-4}$ and $1.3 \times 10^{-4}\,\mathrm{ph\,cm^{-2}\,s^{-1}}$ for $F_{7}$ and $F_{22}$, respectively \cite{Siegert2021_BHMnovae}.
The function $f(f_{\rm Ps},F_{511})$ for the Positronium continuum and 511\,keV line can be found in Ref.\,\cite{Siegert2021_BDHanalysis} and is only parametrized by the Positronium fraction $f_{\rm Ps}$ and the 511\,keV line flux $F_{511}$.
Likewise, the function $g(f_{\rm PBH},M_{\rm PBH})$ is described in detail in the main text and parametrized through the fraction $f_{\rm PBH}$ to account for possible PBH evaporation of a monochromatic PBH mass distribution centered at $M_{\rm PBH}$.

The best-fit spectral parameters are summarized in Tab.\,\ref{tab:fit_parameters}.
The method to derive upper bounds on the PBH fraction vs mass is described in detail in the appendix of Ref.\,\cite{Siegert2021_RetII}.
We refer the reader to this publication.

\begin{table}[]
    \centering
    \begin{tabular}{c|rrr}
        Parameter & Value & Neg. Uncert. & Pos. Uncert. \\
        \hline
        $C_0$ & $3.40$ & $-2.10$ & $1.60$ \\
        $E_C$ & $5.80$ & $-0.60$ & $0.60$ \\
        $C_1$ & $7.00$ & $-0.60$ & $0.60$ \\
        $\alpha_1$ & $-1.24$ & $-0.06$ & $0.06$ \\
        $F_{26}$ & $4.00$ & $-0.50$ & $0.60$ \\
        $b$ & $0.19$ & $-0.07$ & $0.07$ \\
        $F_{7}$ & $1.70$ & $-1.20$ & $1.20$ \\
        $F_{22}$ & $1.10$ & $-0.80$ & $0.80$ \\
        $F_{511}$ & $13.20$ & $-1.60$ & $1.60$ \\
        $f_{\rm Ps}$ & $0.88$ & $-0.09$ & $0.09$\\
        \hline
    \end{tabular}
    \caption{Best fit parameters excluding the PBH component. The units are, from top to bottom, $\mathrm{ph\,cm^{-2}\,s^{-1}\,keV^{-1}}$, keV, $10^{-6}\,\mathrm{ph\,cm^{-2}\,s^{-1}\,keV^{-1}}$, $1$, $10^{-4}\,\mathrm{ph\,cm^{-2}\,s^{-1}}$, $1$, $10^{-4}\,\mathrm{ph\,cm^{-2}\,s^{-1}}$, $10^{-4}\,\mathrm{ph\,cm^{-2}\,s^{-1}}$, $10^{-4}\,\mathrm{ph\,cm^{-2}\,s^{-1}}$, and $1$.}
    \label{tab:fit_parameters}
\end{table}

\bibliography{pbh_spi,thomas}

\end{document}